\documentclass[preprint,journal]{vgtc}       

\ifpdf
  \pdfoutput=1\relax                   
  \usepackage{graphicx}                
  \DeclareGraphicsExtensions{.pdf,.png,.jpg,.jpeg} 
\else
  \usepackage{graphicx}                
  \DeclareGraphicsExtensions{.eps}     
\fi%

\graphicspath{{figures/}{pictures/}{images/}{./}} 

\usepackage{microtype}                 
\PassOptionsToPackage{warn}{textcomp}  
\usepackage{textcomp}                  
\usepackage{mathptmx}                  
\usepackage{times}                     
\usepackage{cite}                      
\usepackage{tabu}                      
\usepackage{booktabs}                  
\usepackage{algorithm}
\usepackage[noend]{algpseudocode}
\usepackage{enumitem}
\usepackage{amssymb}
\usepackage{tikz}
\usepackage{soul}
\usepackage{xcolor}
\usepackage{comment}
\usepackage{tabularx}
\usepackage{xspace}

\usepackage{color}
\newcommand{\name}      {PC-Expo\xspace}
\newcommand{\Cross}     {\mathbin{\tikz [x=1.4ex,y=1.4ex,line width=.2ex] \draw (0,0) -- (1,1) (0,1) -- (1,0);}}

\definecolor{darkgreen}{rgb}{0,0.5,0}
\definecolor{purple}{rgb}{0.75,0,0.75}
\definecolor{brown}{rgb}{0.65,0.16,0.16}
\definecolor{darkslateblue}{rgb}{0.28, 0.24, 0.55}
\definecolor{orange}{rgb}{1.0, 0.647, 0}

\newcount\notenum
\newcommand{\fslnote}[3]        {\global\advance\notenum by 1\textsl{\bf\color{#2}$\blacksquare$[#1\the\notenum: #3]}}

\newcommand{\ie}                {\emph{i.e.},\xspace}
\newcommand{\eg}                {\emph{e.g.},\xspace}
\newcommand{\etal}              {\emph{et~al.}\xspace}

\newcolumntype{C}{>{\centering\arraybackslash}X}

\newcommand{\capfont}                   \small

\onlineid{1631}

\vgtccategory{VIS 2022 Analytics \& Decisions Papers}
\vgtcpapertype{algorithm/technique}

\title{\name: A Metrics-Based Interactive Axes Reordering Method for Parallel Coordinate Displays}

\author{Anjul Tyagi, Tyler Estro, Geoff Kuenning, Erez Zadok, Klaus Mueller}
\authorfooter{
\item
 Anjul Tyagi, Tyler Estro, Erez Zadok and Klaus Mueller are at the Computer Science Department, Stony Brook University, New York. E-mail: \{aktyagi, testro, ezk, mueller\}@cs.stonybrook.edu.
\item Geoff Kuenning is at the Computer Science Department, Harvey Mudd College, Claremont, California. E-mail: geoff@cs.hmc.edu
}

\shortauthortitle{Tyagi \MakeLowercase{\textit{et al.}}: \name\ : A Metrics-Based Interactive Axes Reordering Method for Parallel Coordinate Displays}

\abstract{
Parallel coordinate plots (PCPs) have been widely used for
high-dimensional (HD) data storytelling because they allow for presenting a
large number of dimensions without distortions.
The axes ordering in PCP presents a particular story from the data
based on the user perception of PCP polylines.  Existing works focus
on directly optimizing for PCP axes ordering based on some common
analysis tasks like clustering, neighborhood, and correlation.
However, direct optimization for PCP axes based on these common
properties is restrictive because it does not account for multiple
properties occurring between the axes, and for local properties that
occur in small regions in the data.  Also, many of these techniques do
not support the human-in-the-loop (HIL) paradigm, which is crucial (i) for
explainability and (ii) in cases where no single reordering scheme
fits the users' goals.
To alleviate these problems, we present \name, a real-time visual
analytics framework for all-in-one PCP line pattern
detection and axes reordering.  We studied the connection of line
patterns in PCPs with different data analysis tasks and datasets.
\name expands prior work on PCP axes reordering by developing real-time,
local detection schemes for the 12 most common analysis tasks
(properties).
Users can choose the story they want to present with
PCPs by optimizing directly over their choice of properties.
These properties can be ranked, or combined using
individual weights, creating a custom optimization scheme for
axes reordering.  Users can control the granularity at which
they want to work with their detection scheme in the data, allowing
exploration of local regions.
\name also supports HIL axes reordering via local-property
visualization, which shows the regions of granular activity
for every axis pair.  Local-property visualization is helpful for
PCP axes reordering based on multiple properties, when no single reordering
scheme fits the user goals.  A comprehensive evaluation was done with
real users and diverse datasets confirm the efficacy of
\name in data storytelling with PCPs.
}


\CCScatlist{ 
 \CCScat{K.6.1}{Management of Computing and Information Systems}%
{Project and People Management}{Life Cycle};
 \CCScat{K.7.m}{The Computing Profession}{Miscellaneous}{Ethics}
}

\teaser{
 \centering
 \includegraphics[width=\linewidth]{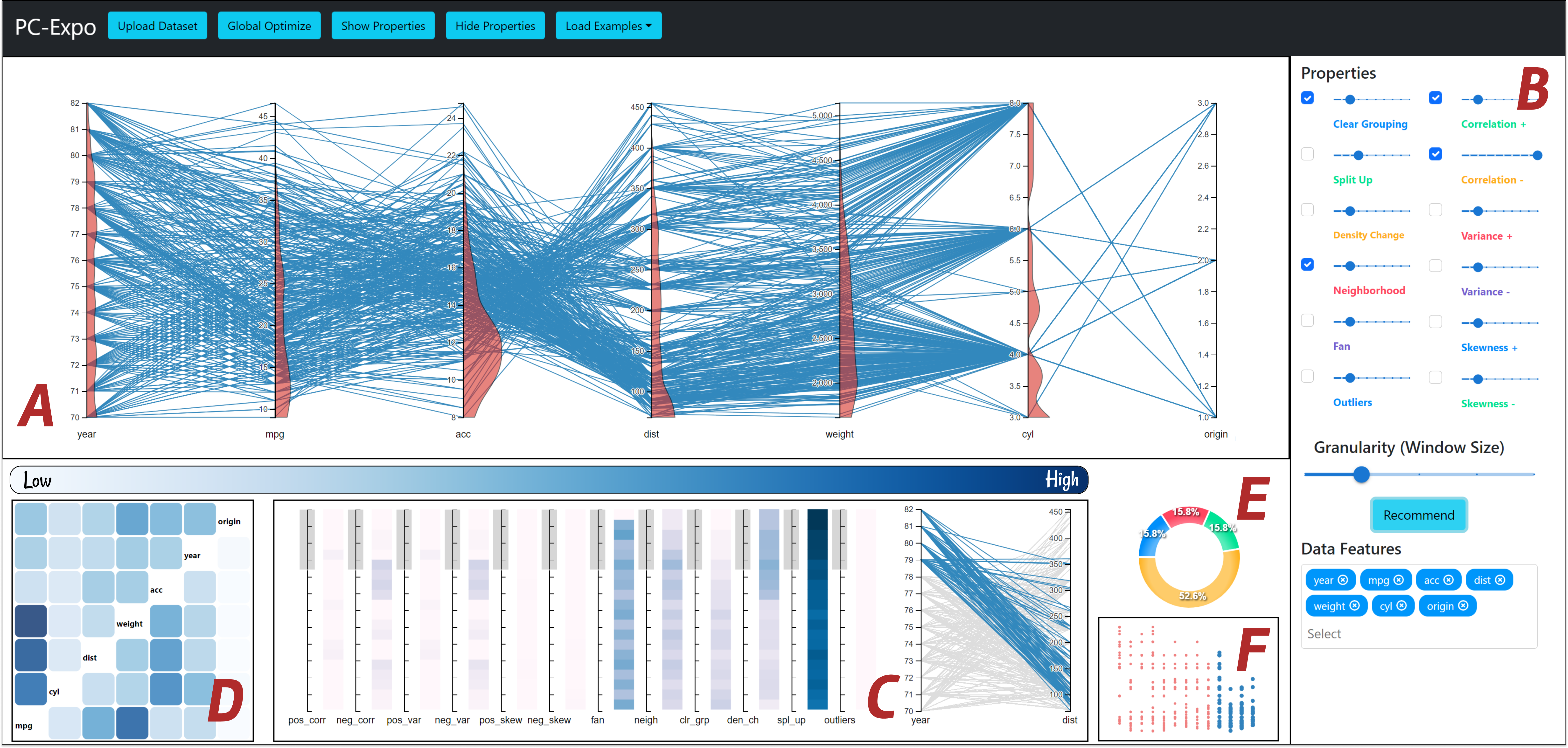}
 \vspace{-6mm}
 \caption{\name is a real-time all-in-one Parallel Coordinate Plot
   (PCP) axes reordering framework.  \name detects local properties in
   high-dimensional data that
   can be used to reorder the PCP axes automatically or
   with human-in-the-loop (HIL) interactions \textbf{\textit{A}}.  We
   have implemented detectors for the 12 most common data properties used
   to reorder PCPs,
   shown on the properties panel \textbf{\textit{B}}.  Users can
   create their own optimization scheme using a weighted sum of these
   properties, by selecting respective properties and weights from (B), summarized as a donut chart \textbf{\textit{E}} for
   automated axes reordering.  \name also supports HIL axes reordering via
   a heatmap and local views \textit{\textbf{D}}, \textit{\textbf{C}}, and
   \textbf{\textit{F}}.  \textit{\textbf{D}} summarizes the weighted sum of
   user-selected properties detected locally for each axis pair.
   \textbf{\textit{C}} shows where these visualization properties were
   detected for a particular axis pair, with a linked scatterplot
   \textbf{\textit{F}} for visualizing the 2D data points.  Users can manually
   reorder the axes using these local views by clicking on
   \textbf{\textit{D}} sequentially.  The granularity slider in
   \textbf{\textit{B}} lets users control the
   size of local regions used to detect the properties.  Area
   charts next to PCP axes in \textit{\textbf{A}} show the local regions
   where the properties selected on (B) are detected on the axis.}
 \label{fig:teaser}
}

\vgtcinsertpkg

\begin{document}



\maketitle
\section{Introduction}

Visual analytics of high-dimensional (HD) data is crucial in many
applications.  Common HD visualization techniques include
embedding methods like MDS~\cite{kruskal_relationship_1977, kumar2019task}
and t-SNE~\cite{van_der_maaten_visualizing_2008}, and  projection
methods like RadViz~\cite{daniels_properties_2012} and Star
Coordinates~\cite{kandogan_star_2000}.  Most of these techniques
suffer from information loss and distortions as the data is
transformed to lower dimensions.
Parallel Coordinate Plots (PCPs)~\cite{inselberg_parallel_1990} have
been a popular choice for HD data visualization since they
can convey a large number of dimensions without distortions.  PCPs are
considered a storytelling method where each ordering of axes presents a particular story from the HD data, and  techniques like axes repetition, data scaling, and axes inversion have
been suggested to enable a persuasive narration of a story~\cite{siirtola2006interacting, heinrich2013state, tyagi2022visualization, tyagi2019ice}.  Previous
storytelling work with PCPs mainly focused on axes arrangement based on
common data properties like correlation, clustering, and the
number of line crossings~\cite{
  tatu_automated_2011}; they detect these properties in every pair of
dimensions in the data and then find the corresponding axes
arrangement using the traveling salesman problem
(TSP)~\cite{flood_traveling-salesman_1956, zhang2012network} over the computed scores.
Yet, given the complex patterns that can appear in HD datasets, there is
no one-solution-fits-all for PCP axes arrangement and storytelling.
Different use cases require conveying different stories through PCPs
and hence require different axes arrangements.

Existing, fully automated
PCP axes-arrangement techniques have four major shortcomings.
The first is the lack of
human-in-the-loop (HIL) support, which limits
their utility to just a few applications.  Second is the lack of capabilities to explore local regions (a subset of records) of the data.  As HD
data are typically complex many properties may only occur
locally.  For example, there can be local regions of positively
correlated clusters in an overall negatively correlated data set.
Depending on the use case, such local clusters could have major
significance.
However, these regions might be completely ignored by fully automated
axes-ordering techniques if they are dominated by other, more global phenomena gauged by a global metric.
While techniques exist that work with local
detection~\cite{makwana_axes_nodate} they
are too slow since they work at a fairly low level of granularity.
In addition, they also lack
adaptability since the size of the local regions cannot be customized.

\begin{figure}[tb]
 \centering
 \includegraphics[width=\columnwidth]{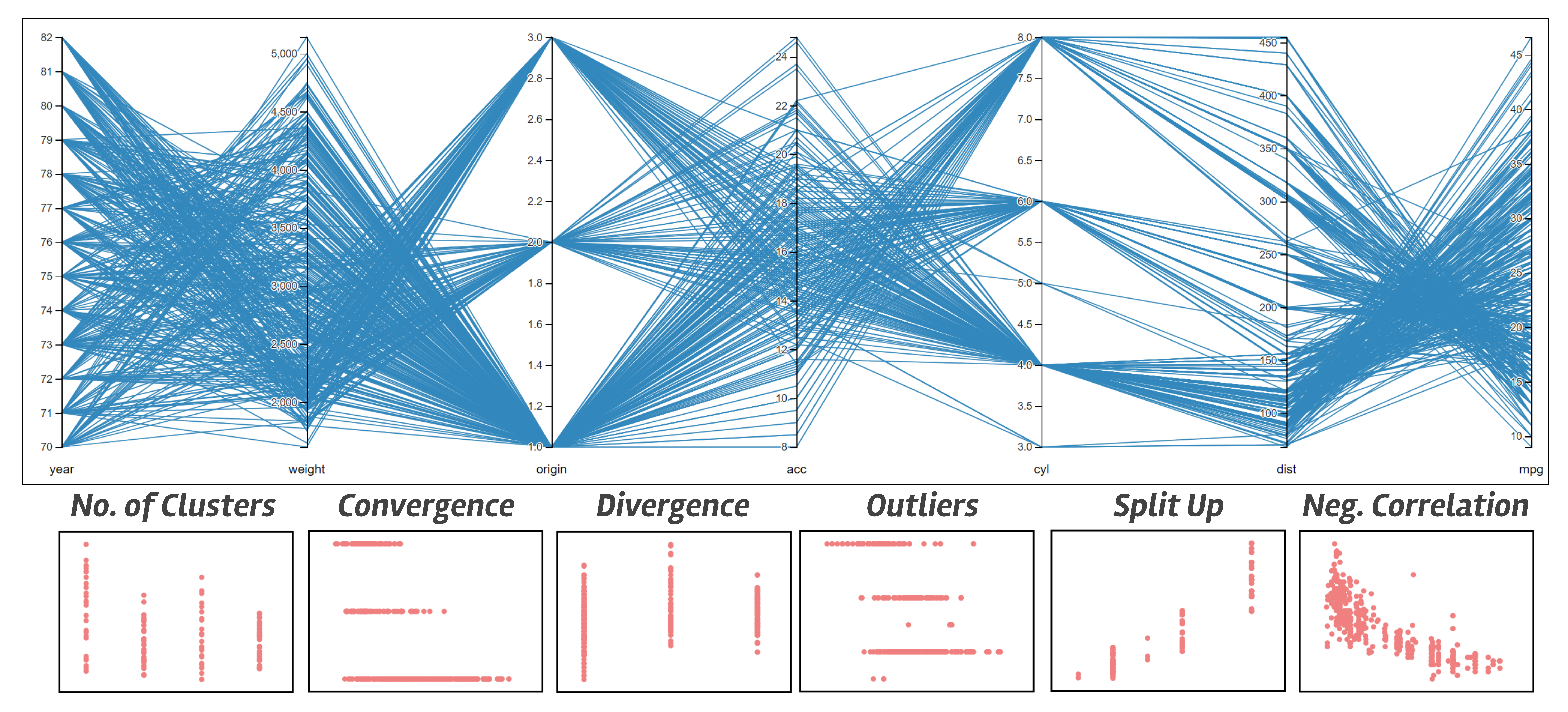}
 \vspace{-6mm}
 \caption{PCP axes reordering based on multiple properties.  In many
   applications, it is desired to reorder PCP axes based on the property
   that pops out the most in a particular axis pair.  The example
   shows axes reordering based on six different properties.
   The scatterplots below each axis pair
   show the data points in 2D.}
 \label{fig:ordering}
\vspace{-10pt}
\end{figure}

Third, there is also a growing interest in
explainability~\cite{xu2019toward, cao2019graphs, tyagi2021user, tyagi2020visual} in modern
computer-human interaction systems.
For visualization frameworks that are
deployed for high-liability applications (\eg crime-fighting),
explainability
is a crucial feature.
Most existing PCP axes-ordering techniques fail
to convey \emph{why} a particular ordering was chosen.
The fourth and final shortcoming involves a lack of support for axes
reordering with multiple patterns.  For example, in
Figure~\ref{fig:ordering} one of the analysts from our user study
preferred to order the PCP axes based on different properties instead
of just using a single property.  This calls for a HIL paradigm and
an all-in-one, explainable PCP storytelling system.

We designed \textit{\name}
(see Figure~\ref{fig:teaser}) to address these pending issues which all call for a more refined
PCP axes reordering methodology.
\name offers real-time detection schemes for
the 12 most common data analysis properties used to reorder PCPs, previously introduced by Blumenschein \etal~\cite{blumenschein_evaluating_2020}. 
Users can create their optimization scheme using these properties
and detect them locally across the data dimensions.  \name supports
automated and manual axes reordering based on these detected properties.
To support a HIL paradigm with explainability, \name incorporates a
%
%
local view of each dimension pair, highlighting regions of local
activity in the data.

Our main contributions (plus code and software release)
\footnote{PC-Expo demo is available at: \url{https://paracoords.herokuapp.com/}}
are:
%
\vspace{-3mm}
\begin{itemize}
\setlength{\itemsep}{-3pt}
\item Localized and real-time detection algorithm implementations for the
  12 most common properties~\cite{blumenschein_evaluating_2020} used for PCP reordering;
\item HIL and explainable PCP axes reordering via local views in \name
  (see Figure~\ref{fig:teaser}, labels C, D, and F);
\item Multi-property axes reordering (see
  Figure~\ref{fig:ordering});
  and
\item A fully automated optimization algorithm for PCP axes reordering based on
  localized property detection.
\end{itemize}


\section{Related Work}

We summarize several lines of PCP axes-reordering research by comparing them
based on the type of properties they can detect and the optimization
schemes they follow (see Table~\ref{tab:comp_hyp}). Based on the 12 most common data analytics properties, that can be used to reorder PCPs as described in ~\cite{blumenschein_evaluating_2020}, Table~\ref{tab:comp_hyp} compares techniques that are used to detect these properties partially. Compared to the individual techniques, \name can be used to detect all the 12 properties available in a single interface.  

\begin{table}[tb]
\caption{Comparing \name with different PCP reordering algorithms
  based on the type of properties (\textbf{Figure~\ref{fig:props}})
  they can detect {\color[HTML]{4472C4}(blue)} and the optimization
  schemes they follow {\color[HTML]{C65911}(brown)}.  \textit{Pairwise
    Axes} optimization refers to techniques optimizing on
  pairwise-dimension values of the detected properties and then
  finding the axes ordering using
  TSP~\cite{dasgupta_pargnostics_2010}.  \textit{Localized axes}
  means that the technique allows local detection of properties in the
  data.
\label{tab:comp_hyp}}
\scriptsize%
\centering%
\begin{tabu}{%
    r%
    *{8}{c}%
    *{16}{r}%
  }
\toprule
\rotatebox{90}{\textbf{Technique}}
& \rotatebox{90}{Makwana} \rotatebox{90}{(Line
  Slopes)~\cite{makwana_axes_nodate}} & \rotatebox{90}{Peng
  (Outliers)~\cite{peng_clutter_2004}} & \rotatebox{90}{Artero}
\rotatebox{90}{(Similarity)~\cite{artero2006enhanced}} &
\rotatebox{90}{LU (SVD)~\cite{lu_two_2016}} & \rotatebox{90}{Ankerst}
\rotatebox{90}{(Correlation)~\cite{berchtold_similarity_1998}} &
\rotatebox{90}{Pargnostics~\cite{dasgupta_pargnostics_2010}} &
\rotatebox{90}{\textbf{Ours}} \\
\midrule
{\color[HTML]{4472C4} \textbf{Clear Grouping}}               &
\color[HTML]{E41A1C} $\Cross$                       &
\color[HTML]{E41A1C} $\Cross$              & \color[HTML]{4DAF4A}
\checkmark                   & \color[HTML]{E41A1C} $\Cross$        &
\color[HTML]{E41A1C} $\Cross$                     &
\color[HTML]{4DAF4A} \checkmark        & \color[HTML]{4DAF4A}
\checkmark   \\
\midrule
{\color[HTML]{4472C4} \textbf{Density Change}}               &
\color[HTML]{E41A1C} $\Cross$                       &
\color[HTML]{E41A1C} $\Cross$               & \color[HTML]{E41A1C}
$\Cross$                   & \color[HTML]{E41A1C} $\Cross$        &
\color[HTML]{E41A1C} $\Cross$                     &
\color[HTML]{4DAF4A} \checkmark        & \color[HTML]{4DAF4A}
\checkmark    \\
\midrule
{\color[HTML]{4472C4} \textbf{Split Up}}                     &
\color[HTML]{E41A1C} $\Cross$                       &
\color[HTML]{E41A1C} $\Cross$               & \color[HTML]{E41A1C}
$\Cross$                   & \color[HTML]{E41A1C} $\Cross$        &
\color[HTML]{E41A1C} $\Cross$                     &
\color[HTML]{4DAF4A} \checkmark        & \color[HTML]{4DAF4A}
\checkmark    \\
\midrule
{\color[HTML]{4472C4} \textbf{Neighborhood}}                 &
\color[HTML]{E41A1C} $\Cross$                       &
\color[HTML]{E41A1C} $\Cross$               & \color[HTML]{E41A1C}
$\Cross$                   & \color[HTML]{E41A1C} $\Cross$        &
\color[HTML]{E41A1C} $\Cross$                     &
\color[HTML]{E41A1C} $\Cross$        & \color[HTML]{4DAF4A} \checkmark
\\
\midrule
{\color[HTML]{4472C4} \textbf{Positive Correlation}}         &
\color[HTML]{E41A1C} $\Cross$                       &
\color[HTML]{E41A1C} $\Cross$               & \color[HTML]{4DAF4A}
\checkmark                   & \color[HTML]{E41A1C} $\Cross$        &
\color[HTML]{E41A1C} $\Cross$                     &
\color[HTML]{4DAF4A} \checkmark        & \color[HTML]{4DAF4A}
\checkmark    \\
\midrule
{\color[HTML]{4472C4} \textbf{Negative Correlation}}         &
\color[HTML]{4DAF4A} \checkmark                       &
\color[HTML]{E41A1C} $\Cross$               & \color[HTML]{E41A1C}
$\Cross$                   & \color[HTML]{E41A1C} $\Cross$        &
\color[HTML]{E41A1C} $\Cross$                     &
\color[HTML]{4DAF4A} \checkmark        & \color[HTML]{4DAF4A}
\checkmark    \\
\midrule
{\color[HTML]{4472C4} \textbf{Positive Variance}}            &
\color[HTML]{E41A1C} $\Cross$                       &
\color[HTML]{E41A1C} $\Cross$               & \color[HTML]{E41A1C}
$\Cross$                   & \color[HTML]{E41A1C} $\Cross$        &
\color[HTML]{4DAF4A} \checkmark                     &
\color[HTML]{4DAF4A} \checkmark        & \color[HTML]{4DAF4A}
\checkmark    \\
\midrule
{\color[HTML]{4472C4} \textbf{Negative Variance}}            &
\color[HTML]{E41A1C} $\Cross$                       &
\color[HTML]{E41A1C} $\Cross$               & \color[HTML]{E41A1C}
$\Cross$                   & \color[HTML]{E41A1C} $\Cross$        &
\color[HTML]{4DAF4A} \checkmark                     &
\color[HTML]{4DAF4A} \checkmark        & \color[HTML]{4DAF4A}
\checkmark    \\
\midrule
{\color[HTML]{4472C4} \textbf{Positive Skewness}}            &
\color[HTML]{E41A1C} $\Cross$                       &
\color[HTML]{E41A1C} $\Cross$               & \color[HTML]{E41A1C}
$\Cross$                   & \color[HTML]{E41A1C} $\Cross$        &
\color[HTML]{4DAF4A} \checkmark                     &
\color[HTML]{E41A1C} $\Cross$        & \color[HTML]{4DAF4A} \checkmark
\\
\midrule
{\color[HTML]{4472C4} \textbf{Negative Skewness}}            &
\color[HTML]{E41A1C} $\Cross$                       &
\color[HTML]{E41A1C} $\Cross$               & \color[HTML]{E41A1C}
$\Cross$                   & \color[HTML]{E41A1C} $\Cross$        &
\color[HTML]{4DAF4A} \checkmark                     &
\color[HTML]{E41A1C} $\Cross$        & \color[HTML]{4DAF4A} \checkmark
\\
\midrule
{\color[HTML]{4472C4} \textbf{Fan}}                          &
\color[HTML]{4DAF4A} \checkmark                       &
\color[HTML]{E41A1C} $\Cross$               & \color[HTML]{E41A1C}
$\Cross$                   & \color[HTML]{E41A1C} $\Cross$        &
\color[HTML]{4DAF4A} \checkmark                     &
\color[HTML]{4DAF4A} \checkmark        & \color[HTML]{4DAF4A}
\checkmark    \\
\midrule
{\color[HTML]{4472C4} \textbf{Outliers}}                     &
\color[HTML]{E41A1C} $\Cross$                       &
\color[HTML]{4DAF4A} \checkmark               & \color[HTML]{E41A1C}
$\Cross$                   & \color[HTML]{E41A1C} $\Cross$        &
\color[HTML]{E41A1C} $\Cross$              & \color[HTML]{E41A1C}
$\Cross$        & \color[HTML]{4DAF4A} \checkmark    \\
\toprule
{\color[HTML]{C65911} \textbf{Pairwise Axis}}       &
\color[HTML]{4DAF4A} \checkmark                       &
\color[HTML]{4DAF4A} \checkmark               & \color[HTML]{4DAF4A}
\checkmark                   & \color[HTML]{E41A1C} $\Cross$        &
\color[HTML]{E41A1C} $\Cross$                     &
\color[HTML]{4DAF4A} \checkmark        & \color[HTML]{4DAF4A}
\checkmark    \\
\midrule
{\color[HTML]{C65911} \textbf{Localized Axes}}      &
\color[HTML]{E41A1C} $\Cross$                       &
\color[HTML]{E41A1C} $\Cross$               & \color[HTML]{E41A1C}
$\Cross$                   & \color[HTML]{E41A1C} $\Cross$        &
\color[HTML]{E41A1C} $\Cross$                     &
\color[HTML]{E41A1C} $\Cross$        & \color[HTML]{4DAF4A} \checkmark
\\
\bottomrule
\end{tabu}%
\end{table}

\begin{figure}[tb]
  \centering
  \includegraphics[width=\columnwidth]{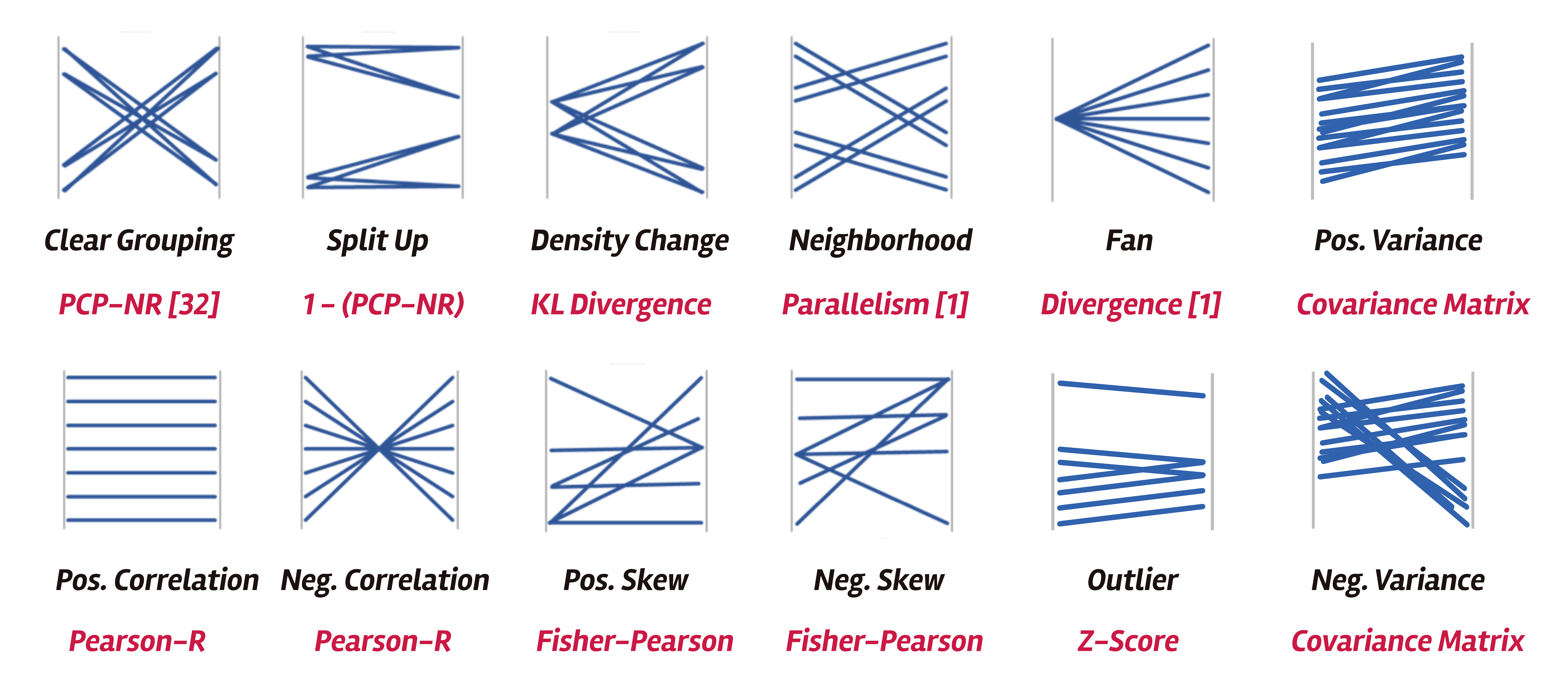}
  \vspace{-6mm}
  \caption{Properties and their corresponding line patterns detectable
    in \name.  Blumenschein \etal~\cite{blumenschein_evaluating_2020}
    proposed these line patterns and related them to common analytical
    tasks.  Below, in red, each analytical task is the method we have
    used in \name to detect the corresponding patterns locally in the
    data.}
 \label{fig:props}
\end{figure}

\subsection{High-Dimensional Data Visualization}

High-dimensional (HD) data come in various types: numerical, ordinal
and nominal.
While some HD data visualization techniques are specifically designed for only a subset of these data types, others apply to all.
%
The most general data type is numerical, and techniques~\cite{zhang2014visual} have been proposed to convert ordinal and nominal into numerical
values for better
visualization and data handling.
A popular paradigm for visualizing HD numerical data is the scatterplot
matrix~\cite{hartigan_printer_1975}, which decomposes the HD space into a set of bivariate projections.  Variants of this approach
include bi-variate projections of the full space and HyperSlices-based
approaches~\cite{berger_uncertainty-aware_2011,
  piringer_hypermoval_2010}.
However, these techniques do not scale with the number of attributes as the
number of plots increases exponentially.  Even though quality metrics like
DSC~\cite{sips2009selecting} can be used for automated plot selection and reduction, it remains difficult to mentally fuse the disjoint
relationships conveyed in the individual plots.

%

Techniques that use non-linear optimization to embed the
data into a 2D plane can overcome these challenges.  They include
MDS~\cite{kruskal_relationship_1977}, Kernel PCA, Local Linear
Embedding~\cite{roweis_nonlinear_2000}, Spectral
Clustering~\cite{ng_spectral_2001}, and
t-SNE~\cite{van_der_maaten_visualizing_2008}.
%
However, these techniques lose the context to the attributes in the
data transformation process, making it difficult for users to discern the
semantics of the embedded patterns.  Conversely, linear
transformations into a radial layout as produced by
RadViz~\cite{daniels_properties_2012, grinstein_high-dimensional_2001,
  hoffman_dna_1997} and Star Coordinates~\cite{kandogan_star_2000,
lehmann_optimal_2016} preserve this context, but they cannot preserve
distance relationships which can lead to ambiguities in the
layout.  Adding an optimization step can recover these
relationships~\cite{cheng2017radviz}.


PCPs have the unique ability to visualize HD data without any
distortions or ambiguities.  Their inherent shortcoming is that the
information that can be visually extracted is dependent on the axes
ordering.  Scalability to large datasets is also a problem as the
increasing number of polylines leads to cluttered displays which are
hard to read.
To alleviate the latter, Ellis and Dix~\cite{ellis_enabling_2006}
calculate the over-plotting percentage
in PCPs and provide a lens system to see through highly cluttered
regions in the plot.
An improvement of this work includes clutter-reduction techniques in
PCPs~\cite{ellis_taxonomy_2007}.  Illustrative
PCP~\cite{mcdonnell_illustrative_2008} aims to reveal different
patterns in PCPs lines using edge bundling, axes distribution, and
cluster visualization.
Johansson \etal~\cite{johansson_revealing_2005} discuss four methods
to reveal clusters in PCPs using data-transformation techniques.  They
propose that data scaling using root and log transforms can reveal
patterns in PCP lines.
Pomerenke \etal~\cite{pomerenke_slope-dependent_2019} describe the
existence of ghost clusters in PCPs because of line patterns, which
can be prevented by adjusting line widths in PCPs at specific
positions based on projection angles.
Peng \etal~\cite{peng_clutter_2004} show how to properly cluster the
PCP lines with a strategy that assigns each point to a PCP cluster.
Improving upon clustering with PCPs, Peltonen
\etal~\cite{peltonen_parallel_2017} describe how to find local and
global neighbors of data points in a PCP.

All of these techniques focus on detecting patterns in an
\textit{already existing} PCP representation.  However, as mentioned,
the axes ordering plays a crucial role in generating the data pattern
that are actually revealed to the user and the detector.  In \name, we
take one step back and allow users to prioritize and detect the
patterns in the generation stage which greatly increases the yield.
It allows users to reorder the PCP's axes to
present the desired pattern-based story through PCPs.


\subsection{HD Data Visualization Metrics}
\label{ss:rw_hd_data_vis}

HD data metrics are used to bridge the gap between visual
representation through charts and analytical tasks that are commonly
performed on such data.  Visualizations are affected by the choice of
metrics that analysts use to create them; this resonates with the idea
that every visualization has a story to tell, and that these metrics
help in quantifying how well a visualization depicts the
story~\cite{miller_need_1997}.
Tufte~\cite{tufte_2001} was the first to propose quality
metrics for 2D charts, which spawned several additions to
better represent information in 2D scatterplots.
Eagan \etal~\cite{eagan_low-level_2005} relates these metrics to most
%
%
common data-analytics tasks, listing correlation and cluster
identification as the two most important quality metrics for
scatterplots.
Bertini \etal~\cite{bertini_by_2004} suggests sampling metrics for
scatterplots to reduce the cluttering of points.
Scagnostics~\cite{tukey_computing_1985} suggests metrics for detecting
visual structures in scatterplots, which were extended by Wilkinson
\etal~\cite{wilkinson_graph-theoretic_2005}.

Along with metrics for 2D data, specific metrics have been devised
for analyzing line patterns in PCPs.
Blumenschein
\etal~\cite{blumenschein_evaluating_2020} suggest 12 metrics and their
line patterns in PCPs which can impact the quality and visual
representation of the plots.
Pargnostics by Dasgupta \etal~\cite{dasgupta_pargnostics_2010}
%
%
proposed 6 metrics for
PCPs and a reordering strategy based on these properties.
%
Users could create a weighted optimization scheme based on the
Pargnostics
%
%
properties, which motivated our development of \name.
Beyond Pargnostics, other metrics covered in
Blumenschein are convergence of lines~\cite{makwana_axes_nodate},
outliers~\cite{peng_clutter_2004}, skewness and
variance~\cite{lu_two_2016}, and axes
similarity~\cite{berchtold_similarity_1998}.  These metrics can assist
in storytelling with PCPs.  However, they are not available in a
single interface and some of these techniques are unusably slow on large
datasets, as seen in our evaluation results (Section~\ref{s:evaluation}).
\name extends these properties by adding PCP axes reordering with
positive and negative skewness and variance (Figure~\ref{fig:props}).
We developed real-time detectors for these
line patterns using analytical methods and parallel programming, which
are all available to users in a single interface.  Also, users can
detect these line patterns locally in the data, a feature that is not supported
in any prior work that we are aware of.

\subsection{Axes Reordering in Parallel Coordinate Plots}

Axes reordering in PCPs play a crucial role in presenting desired
information accurately.  This has been a major application for the metrics devised
for PCPs (see Section~\ref{ss:rw_hd_data_vis}).
Some of the algorithms optimize to find the ordering directly on the
data~\cite{artero2006enhanced, johansson_evaluation_2015} while others
calculate the PCP metrics for every axis pair before
optimizing~\cite{dasgupta_pargnostics_2010, makwana_axes_nodate,
  peng_clutter_2004}.
After the metrics are calculated for every axis pair, TSP solvers can
be used to find the PCP axes reordering from the data~\cite{dasgupta_pargnostics_2010}.
%
A major limitation of
existing algorithms is the lack of localized detection of these
metrics.
For example, there can be local regions of positively
correlated clusters in an overall negatively correlated data set.
%
%
%
With metric detection without localized support, the algorithm
will return zero correlation for such data.
With existing techniques, such local attributes
in the data go undetected during PCP reordering.
Also, it is hard for the users to clearly visualize why a particular
ordering was chosen, and there is no way to enter user feedback into
these algorithms.  In \name, we introduce localized detection of all
the line patterns in PCPs; and these can be fully controlled by users.
Users can choose the localization level, or even reorder the axes
based on different metrics manually, using the scores obtained via our
detection scheme.


\section{Formative Study}
\label{s:formative_study}

To systematically evolve our idea of an all-in-one PCP axes-ordering
framework, we applied Munzner's nested
model~\cite{munzner_nested_2009} for visualization application design.
Before designing \name, we conducted a formative study to get to know
user requirements and their views on the state-of-the-art
PCP tools, libraries, and general workflows.  This approach helped us
concretize \name's design with a user-centered evaluation at an
earlier development stage.

The formative study participants were carefully chosen to be analysts
and researchers who use PCPs as their regular visualization technique
for HD data.  A total of 10 participants contributed to this study,
out of which 4 were visualization researchers (3 Ph.D.\ students and 1
professor), 3 systems researchers (1 Ph.D.\ student and 2 professors),
and 3 data analysts working in the industry.  While the data analysts
and visualization researchers helped us design the principles related
to visual aspects of PCPs, the systems researchers helped us make the
PCPs more generic and applicable across different domains.
%
%
One of the major contributions from systems researchers
%
%
in this aspect was in optimizing the detection schemes and extending
their usability with parallel programming and low-level
optimizations.

We interviewed our participants in several meetings, asking
questions about the pros and cons of the existing techniques and
tools they use to design PCPs, getting suggestions for improving the current
systems, and asking about any important additions they believed would
be useful in a new system.  All of these interviews and answers were used to
iteratively design \name's final form.

\subsection{Key Findings}

The purpose of performing the formative study with domain experts and
potential users were to gather a list of
requirements that we
expected our framework to meet.  Our many discussions culminated
in the following requirements list, ordered by importance:
%
%

\textbf{R1: Localization.}
None of the existing PCP ordering algorithms show where the properties
lie on the axes.  Highlighting these properties on the PCP axes helps
the user understand how the final ordering was obtained and what
portion of the data contributed to the final ordering.

\textbf{R2: Real-time metrics calculation.}
Existing PCP reordering techniques either (i) work for only a few metrics
or (ii) are more generic but are too
slow~\cite{dasgupta_pargnostics_2010}, as seen with the
%
%
existing tools\footnote{PCP Reordering Implementation
  \url{https://subspace.dbvis.de/pcp/}}.
Thus, a more flexible and faster implementation for real-time
detection would be useful for PCP axes ordering applications.

\textbf{R3: HIL axes reordering.}
\name should support HIL axes reordering because no single
optimization scheme can fulfill all use cases.  For flexibility, users
should be able to use their domain knowledge and feedback from our
local detector algorithms to reorder the PCP axes.

\textbf{R4: Aggregate view.}
The aggregate view has two uses.  For local HIL axes reordering,
\name should quantify each of the properties selected by the users
on all the axis pairs in the data.
%
For automated axes ordering, users should be able to visualize the
weighted properties and the corresponding optimization function on
the interface.

\textbf{R5: Multi-property axes reordering.}
As shown in Figure~\ref{fig:ordering}, users sometimes need to investigate
different patterns in their data, which cannot be accomplished using a
single reordering strategy.  \name should support multi-property axes
reordering for such use cases.


\section{Properties in \name}

Figure~\ref{fig:props} shows the 12 properties supported in \name and
their corresponding
line patterns.  Users can create a custom
optimization scheme using a weighted sum of these properties, each of which
our system then detects in the data.  We
implemented individual detectors that look for
local regions in the data using a sliding window of a user-specified
size, and then calculate the corresponding property values at each window position.
Each property is calculated for every primary (left) and secondary
(right) axis of an axis pair.
In the descriptions below we refer to the primary axis as $X$ and
the secondary axis as $Y$.
Out of the 12 properties, 7 are calculated
directly from the 2D bivariate data defined by the axis pair, \ie
\textit{correlation, variance, density change,
clear grouping, split up, neighborhood} and \textit{fan};
the other 5 are calculated on the marginal axis (\ie the primary axis)
only.  We describe the detection algorithms for each
property next.

\subsection{Correlation, Variance, Skewness, and Outliers}

To find correlation, we compute the Pearson correlation
directly using Equation~\ref{eq:corr};  $r_{XY}$ is
the correlation coefficient calculated using every point $x_i;i=1...N$ and  $y_i;i=1...N$.
Here, $x_i;i=1...N$ is the set of points falling within the
sliding window on the primary axis $X$.
The points $y_i;i=1...N$ are the
corresponding tuple values of $x_i;i=1...N$ on the secondary axis $Y$.
$\bar{x}$ and $\bar{y}$ are the mean values of all $x_i$ and $y_i$,
respectively.
The variance can be directly inferred as the numerator value in
Equation~\ref{eq:corr}.

\begin{equation}
r_{XY} = \frac{\sum_{i=1}^{N} (x_{i} - \bar{x}) (y_{i} - \bar{y})}{\sqrt{\sum_{i=1}^{N}(x_{i}-\bar{x})^2} \sqrt{\sum_{i=1}^{N}(y_{i}-\bar{y})^2}}
\label{eq:corr}
\end{equation}

Skewness is a marginal property in \name with values in $[-1,1]$ to indicate negative and positive skew.
It is calculated using the Fisher-Pearson coefficient $g$
shown in Equation~\ref{eq:skewness}, where $m_{i}$ is the $i_{th}$
central moment of the data.

%
%

\begin{equation}
g=\frac{m_{3}}{m_{2}^{3 / 2}}=\frac{\frac{1}{N} \sum_{i=1}^{N}\left(x_{i}-\bar{x}\right)^{3}}{\left[\frac{1}{N} \sum_{i=1}^{N}\left(x_{i}-\bar{x}\right)^{2}\right]^{3 / 2}}
\label{eq:skewness}
\end{equation}






Finally, the outliers calculation within the given window is done by
calculating the points falling outside the range of $1.5*(Q_1 - Q_3)$
where $Q_1$ and $Q_3$ are the first and third quartile, respectively,
and marking them as outliers.  The total number of outlier points is
then the outlier score for that sliding window.

\subsection{Density Change}

In PCPs, the density of lines between any pair of axes in the data can
change.
%
%
In a local context, this is the difference between
densities of the points on a pair of axes for a given window.
For quantifying the change in densities of two marginal distributions,
we adopted the idea of \textit{Kullback-Leibler (KL)} divergence,
which has been widely used in data-analytics applications.
%
\textit{KL} divergence gives a score based on the difference in
the entropies of two probability distributions.  For our purposes, the first
step is to estimate the probability density functions (\textit{pdf})
of the points on the two marginal axes, $X$ and $Y$.  We use \textit{Kernel Density
  Estimation}~\cite{kish1965survey, silverman2018density} with
a Gaussian kernel to estimate the probability densities of
the data points, shown in Equation~\ref{eq:kde}.

\begin{equation}
p_K(x') = \sum_{i=1}^{N} K(x'-x_i;h)
\label{eq:kde}
\end{equation}

\noindent
where $p_K(x')$ is the estimated pdf for point $x'$ within a group of
points $x_i;i=1...N$ on the $X$ axis, which is estimated using the Gaussian kernel $K$
with a bandwidth parameter $h$ (see Equation~\ref{eq:kernel}).  The
bandwidth parameter controls the perimeter of points around $x'$ that
are to be considered for estimating the density.  For our use case, we
set $h$ to be equal to the standard deviation of the points in the
current window, to speed up calculations and improve result
consistency.

\begin{equation}
K(x';h) \propto \exp(-\frac{x'^2}{2h^2})
\label{eq:kernel}
\end{equation}

After the densities for each point in a window are estimated, we can
compute the $KL$ divergence between the two axes using
Equation~\ref{eq:kld}.

\begin{equation}
D_{KL}(X || Y) = \sum_{i=1}^{N} p_K(x_{i}) log(\frac{p_K(x_{i})}{p_K(y_{i})})
\label{eq:kld}
\end{equation}


\subsection{Clear Grouping and Split-Up}

In PCPs, clear grouping and split-up are inverse properties.  Clear
grouping refers to points that occur in a cluster on one axis and are
also clustered on another axis.  Conversely, split up means that
clustered points on one axis are further apart on the other.
To calculate the value of clear grouping, we use the idea of local
neighbors on a marginal axis as proposed by Peltonen
\etal~\cite{peltonen_parallel_2017}.  Shown in
Equation~\ref{eq:neigh_prob}, for a given point $x'$ on a set of local
points $x_i;i=1...N$ on the axis $X$, the probability that point $x''$ is the
neighbor of point $x'$ can be estimated as a probability distribution
function.  Equation~\ref{eq:neigh_prob} estimates the probability of point $x'$ being a
neighbor of point $x''$.

\begin{equation}
p_X(x'|x'') = \frac{\exp(-(x''- x')^2/\sigma_{X}^2)}{\sum_{x_{i} \neq x''} \exp(-(x''- x_{i})^2/\sigma_{X}^2)}
\label{eq:neigh_prob}
\end{equation}

While comparing points from the two axes, we estimate the neighborhood
probabilities of all points compared to the other points on the axes
using Equation~\ref{eq:neigh_prob}.
%
%
This can be treated as a pdf of neighborhood probabilities for each
point.  The KL divergence between $p_X(x'|x'')$ and $p_{Y}(y'|y'')$ for a
point $(x',y')$ on two axes $X$ and $Y$ shows the number of neighbors that
are still intact when transitioning from axis $X$ to $Y$ (see
Equation~\ref{kld_neigh}).
%

\begin{equation}
D_{KL}(p_X^{x'}, p_{Y}^{y'}) = \sum_{x' \neq x''} p_X(x'|x'') log (\frac{p_X(x'|x'')}{p_{Y}(y'|y'')})
\label{kld_neigh}
\end{equation}

To extend Equation~\ref{kld_neigh} to all the points in a local
window, we can sum the values for all points; this gives us an
estimate of grouping behavior on a pair of axes across all points, as
shown in Equation~\ref{eq:kld_cg}.
For consistent scoring, we have to limit the $D_{X,Y}$ values to a fixed range across the dataset.
However, the range of $D_{X,Y}$ depends on the input data; hence we
normalize $D_{X,Y}$ based on the
precomputed \textit{mean} and \textit{standard deviation} of a sample of $D_{X,Y}$ values from the dataset
to produce $D_{X,Y}(norm)$.
Finally, for split-up calculations, since we define split up as an inverse property of
clear grouping, we obtain split-up values using $1-D_{X,Y}(norm)$.

\begin{equation}
D_{X,Y} = \sum_{i} D_{KL}(p_X^{x_i}, p_{Y}^{y_i}) = \sum_{i} \sum_{j \neq i} p_X(x_j|x_i) log (\frac{p_X(x_j|x_i)}{p_{Y'}(y_j|y_i)})
\label{eq:kld_cg}
\end{equation}

%

\subsection{Neighborhood and Fan}

In the context of PCP line pattern detection, \textit{neighborhood}
refers to the amount of parallelism in lines and \textit{fan} refers
to the divergence of points that originate from a small region on the
left axes.  We calculate these values from the existing PCP
metrics proposed in Pargnostics~\cite{dasgupta_pargnostics_2010} as
\textit{parallelism} and \textit{divergence}, respectively.
%
\textit{Parallelism} is calculated using the extent of the angle
distribution of lines for an axis pair, because parallel lines tend to
have a lower range of line angles that occur between an axis pair.
Similarly, divergence is calculated using a 2D histogram, counting the
number of bins with a value greater than zero on the secondary axis for a
given bin on the main axis, respectively.

\subsection{Confidence Scores and Normalization}

When dealing with the calculation of metrics locally, it is crucial to
normalize the scores based on the global properties of the axes for
consistency and misinformation prevention.  We developed normalization
schemes for each of the 12 properties in \name to ensure that the
calculated values are presented based on their confidence scores.  For
\textit{correlation} and \textit{variance} reporting, the final scores
are normalized based on the p-values, $p_{r}$ obtained for a correlation value, $r_{XY}$ (see Equation~\ref{eq:corr}) for a given
pair of axes $X$ and $Y$.
Referring to $r_{XY}$ as $r$, the $p_{r}$ calculation is shown in Equation~\ref{eq:corr_p} ($N$ is the number points under the sliding window).

\begin{equation}
p_{r} = \frac{r\sqrt{N-2}}{\sqrt{1-r^2}}
\label{eq:corr_p}
\end{equation}

Similarly, for \textit{skewness}, the p-values
used for normalization are
calculated using permutation tests. Also, in case of negative values for correlation, variance, and skewness, the values are inverted to represent higher scores when the actual values are lower.
%
For KL divergence-based properties---\textit{density change, split-up
  and clear grouping}---no normalization is required since KL
divergence automatically accounts for the number of samples used in the
calculation.
Hence, the values are consistent across dimensions.  For
the Pargnostics-based properties~\cite{dasgupta_pargnostics_2010}
\textit{neighborhood} and \textit{fan} we scale the
values based on the fraction of points they were calculated for.
Hence, for a local window with fewer points, a large value is scaled
down, as compared to a window with more points.  Finally, no
normalization is required for outliers as they are calculated directly
over the full range of axis pairs.


\section{\name Design}

To implement our idea of a real-time, HIL, explainable and localized
PCP axes-reordering framework, we developed \name with the help of the
principles discovered during the formative study discussed in
Section~\ref{s:formative_study}.  \name allows users to interactively
visualize and optimize the PCPs on their data and convey the desired story
with high accuracy and confidence.  As shown
in Figure~\ref{fig:teaser}, \name consists of six views, which we
discuss next.  The \textit{R\#} next to each view denotes the
requirements from the formative study that the particular view satisfies.

\subsection{PCP Display with Localized Area Charts (R1-R5)}

\begin{figure}[tb]
 \centering
 \includegraphics[width=\columnwidth]{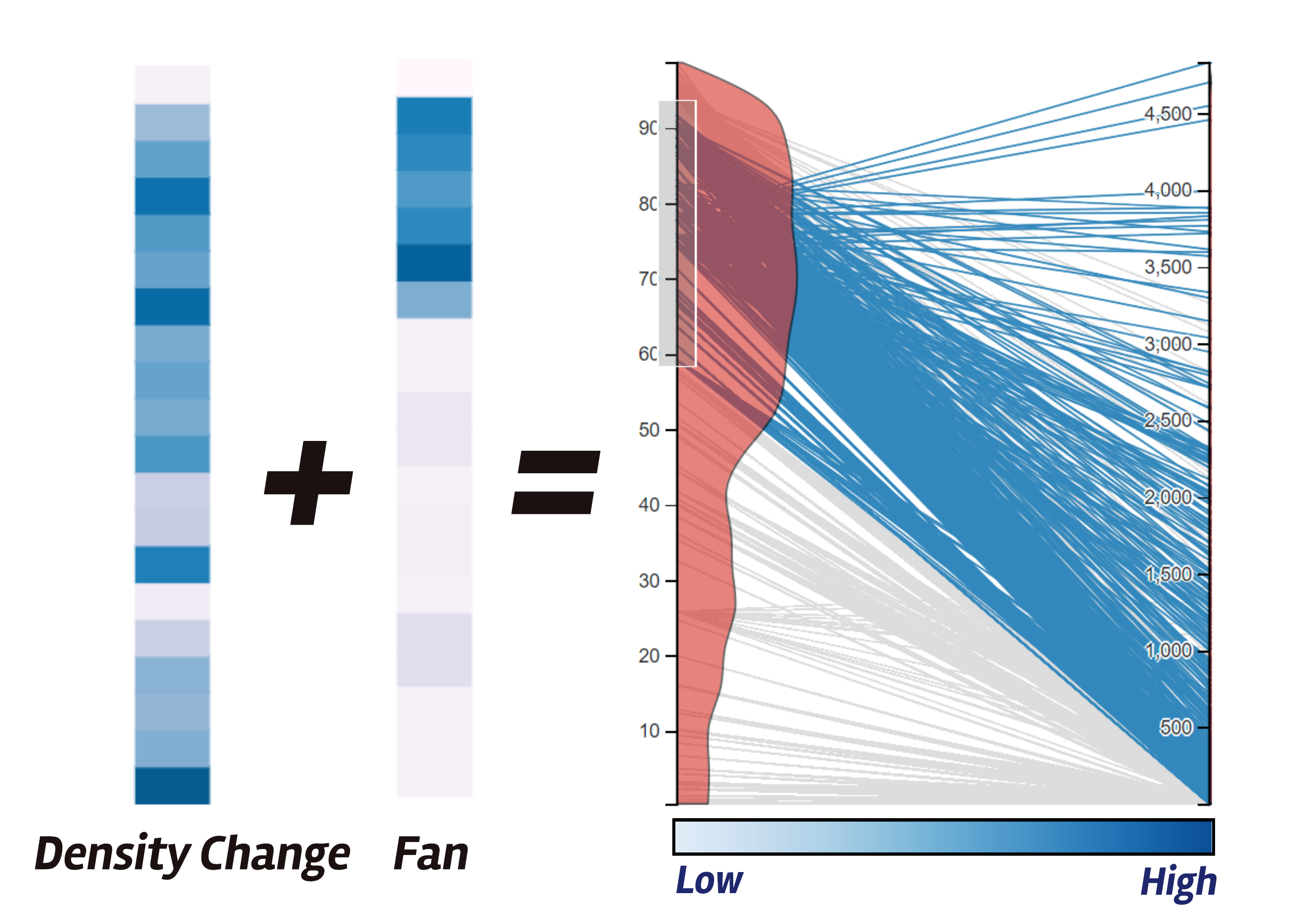}
 \vspace{-6mm}
 \caption{Property summarization view with area charts
   overlaid on the PCP axes in Figure~\ref{fig:teaser} (A).  This
   example shows the user optimizing over the \textit{density change} and
   \textit{fan} properties, which are detected locally with
   intensities shown as heatmaps corresponding to the left axis of the
   PCP.  These intensities are summarized in the form of area charts,
   showing the distribution of where the properties were detected in the
   data.}
 \label{fig:props_sum}
\end{figure}

Shown in Figure~\ref{fig:teaser} (A), the goal of the PCP display view
is to show the final PCP resulting from the user interactions or
direct optimizations.  This view summarizes how the properties chosen
by the user affect the axes reordering with the help of area charts
overlaid on the PCP axes.  As shown in Figure~\ref{fig:props_sum},
this aggregate view helps visualize where the user-selected
properties were detected locally in the final PCP; this is crucial for
explainability of our optimization algorithm's final result.
Users can choose the granularity of the local regions to explore and
look for line patterns; this updates the granularity of the local heatmaps
and the area chart in this view.

\begin{figure*}[!ht]
 \centering
 \includegraphics[width=\textwidth]{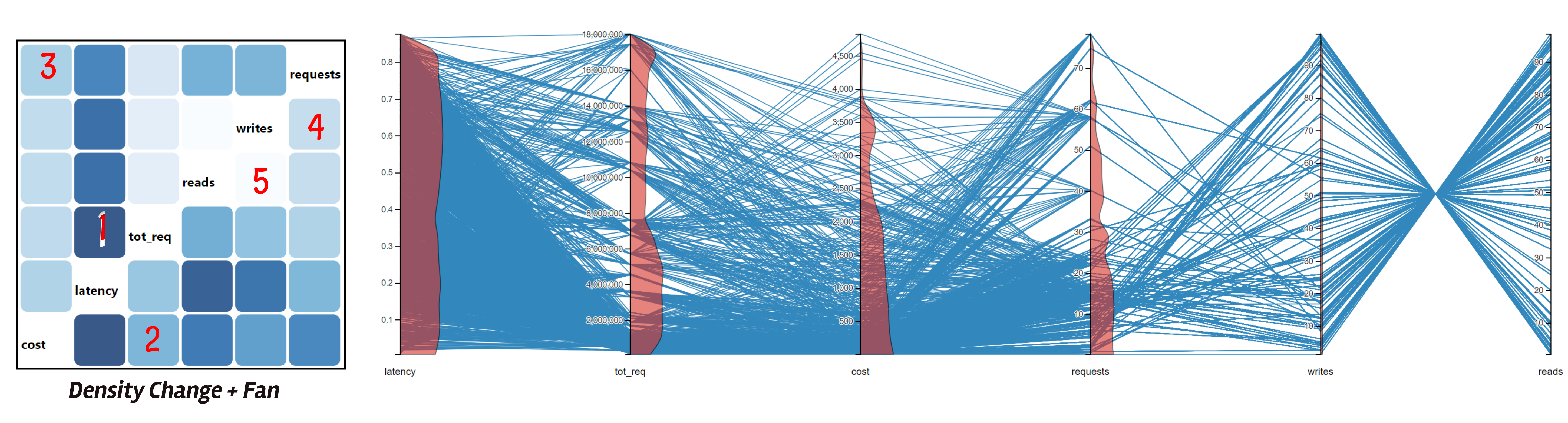}
 \vspace{-8mm}
 \caption{
   HIL axes reordering using a heatmap on \name with decreasing
   scores of the \textit{Density Change} \& \textit{Fan} properties.  The
   heatmap summarizes the weighted sum of properties for
   every axis pair in the data.  Users can double-click on each cell
   in a sequence to reorder the PCP based on the their goals.  In this
   case, the numbers 1--5 show the click sequence on the heatmap to
   generate the PCP displayed on the right.}
 \label{fig:heatmap}
\end{figure*}

\subsection{Sidebar for Local Optimization (R1, R2, R5)}

Shown in Figure~\ref{fig:teaser} (B), the sidebar offers full control
to the users to reorder the PCP axes based on the properties they like.
Users can choose from the 12 properties (see Figure~\ref{fig:props})
with adjoining checkboxes; further, users can choose weights for each
property using the sliders.  These weights are useful when optimizing
directly over these properties to optimize a PCP axis' order: they let
users control how much contribution from each property should be
considered for a final optimization.
For direct optimization, \name calculates the weighted sum of these
property values for each axis pair in the data, where the weights are
determined by the slider values.  Once final scores for each
axis pair are calculated based on the localization level chosen by the
user, a TSP~\cite{dasgupta_pargnostics_2010} over these scores gives the final PCP axes ordering.
The sidebar also offers localization control with the
\textit{window size} slider, which controls the length of the sliding
window.  The value on the \textit{window size} slider is the percent
of axis range to be considered for localized properties detection on
the axis.  Users can also filter the features they want to use in the
final PCP using the \textit{data features} selection tab.


\subsection{Local View (R1, R3, R5)}

Shown in Figure~\ref{fig:teaser} (C, F), the local view is the key
component of \name for allowing HIL and explainable PCP axes reordering.
Based on the property weights and the localization level (granularity)
chosen by the users
in the sidebar, our algorithm detects local regions of activity in the
data.
The local view communicates these local
regions to the user, assisting
%
%
in decision making and understanding the data behavior.
Any axis pair in the data can be selected from the heatmap, as seen in
Figure~\ref{fig:teaser} (D); the local view shows the details of all
properties and their intensities on the chosen axis pair.

The local view also supports user interactions to enable a deeper dive
into local data regions.  Each property is given its individual axis
with a fixed slider width; the sliders can be dragged to highlight the
corresponding points on the adjoining PCP and scatterplot displays, as
seen in Figure~\ref{fig:teaser} (F).  This helps users identify
small regions of interest based on specific properties in the data.

\subsection{Heatmap for HIL Axes Reordering (R3, R4, R5)}

Shown in Figure~\ref{fig:teaser} (D), the heatmap summarizes the
scores from the detection algorithms based on the properties chosen by
the user.  Each cell in the heatmap corresponds to the value of
normalized weighted sum of properties detected for a pair of
dimensions in the dataset.
%
%
Clicking on any cell opens up a window with details of the local view,
showing how the final score for the current cell was calculated.
One major benefit of the heatmap is to assist in HIL axes reordering
based on the calculated scores from the detectors.  As shown in
Figure~\ref{fig:heatmap}, users can view the details of each cell in
the heatmap and use that information to reorder the PCP axes.  Also,
this reordering does not have to stick to a single property and can
include multiple properties per axis pair, as shown in
Figure~\ref{fig:ordering}.  Users can generate a new heatmap after
every axis-pair selection to select the next adjoining axes, allowing
for PCP reordering based on multiple properties.
Each property score is
normalized between [0,1] before calculating the weighted sum values
for the heatmap.


\subsection{Property Weights and Global Optimization (R2)}

For quick results, \name supports a fully automated global
optimization scheme.  With the help of our global optimization
algorithm, users can generate a PCP axes reordering based on selected
properties, corresponding weights, and the localization level.  This
algorithm performs a branch-and-bound TSP algorithm~\cite{dasgupta_pargnostics_2010}
on the
calculated heatmap data and generates PCP axes reordering automatically.
The optimization function showing each property's contribution to the
axes reordering is summarized using a donut chart; see
Figure~\ref{fig:teaser} (E).



\section{Evaluation}
\label{s:evaluation}

This section presents an evaluation of \name for its effectiveness and design
efficiency, through a comparison with existing PCP reordering tools, a
system usability study~\cite{bangor2009determining} survey, and
detailed user interviews.

\subsection{User Study}

We conducted a user study to evaluate \name for its ability to support
PCP axes reordering tasks with real users.  The study aimed at
investigating \name's support as an interactive visual-analytics
system and quantifying how well the system meets user needs.  We
designed three tasks for this study to compare \name with the existing
baseline PCP axes ordering algorithms presented in
Table~\ref{tab:comp_hyp}.  All these baseline algorithms were compiled
into a single dashboard by Blumenschein
\etal~\cite{blumenschein_evaluating_2020}, known as the
\textit{dimensional reordering for parallel coordinates (DRPC)}.  We
carefully designed the tasks for this user study so they can be
performed with both \name and DRPC.  Since \name supports more
properties and a detailed local view---not available in DRPC---we
evaluated these features separately following up from this user study
(see Sections~\ref{ss:sus} and~\ref{ss:user_interviews}). Since \name gives a detailed view of every axis pair weighted sum while performing the reordering tasks, users could memorize the correct sequence for the task. To prevent this, all the user study tasks were first performed on DRPC, followed by \name.

\subsubsection{Operation Details}
DRPC and \name support different types of operations which were noted during the user study. There are a total of 22 operations supported in DRPC which include choosing between 7 reordering strategies, 2 types of distance metrics, and 6 weights-specific Pargnostics PCP reordering. \name supports multiple operations which can linearly increase with data dimensionality. Besides fixed operations for choosing the properties and local view interactions, the matrix view and the main PCP view depend on the data dimensionality.

\subsubsection{Participants}

We recruited 10 participants for this study (4 females and 6 males;
aged between 22 and 30 years) via social media and mailing lists.
5 experts were the same participants from the formative study (Sec~\ref{s:formative_study}. We recruited 5 additional non-experts to participate in this study. All of the non-experts were graduate
students studying computer science.

\subsubsection{User Study Tasks}
\label{sss:user_study_tasks}

\begin{figure}[!ht]
 \centering
 \includegraphics[width=\columnwidth]{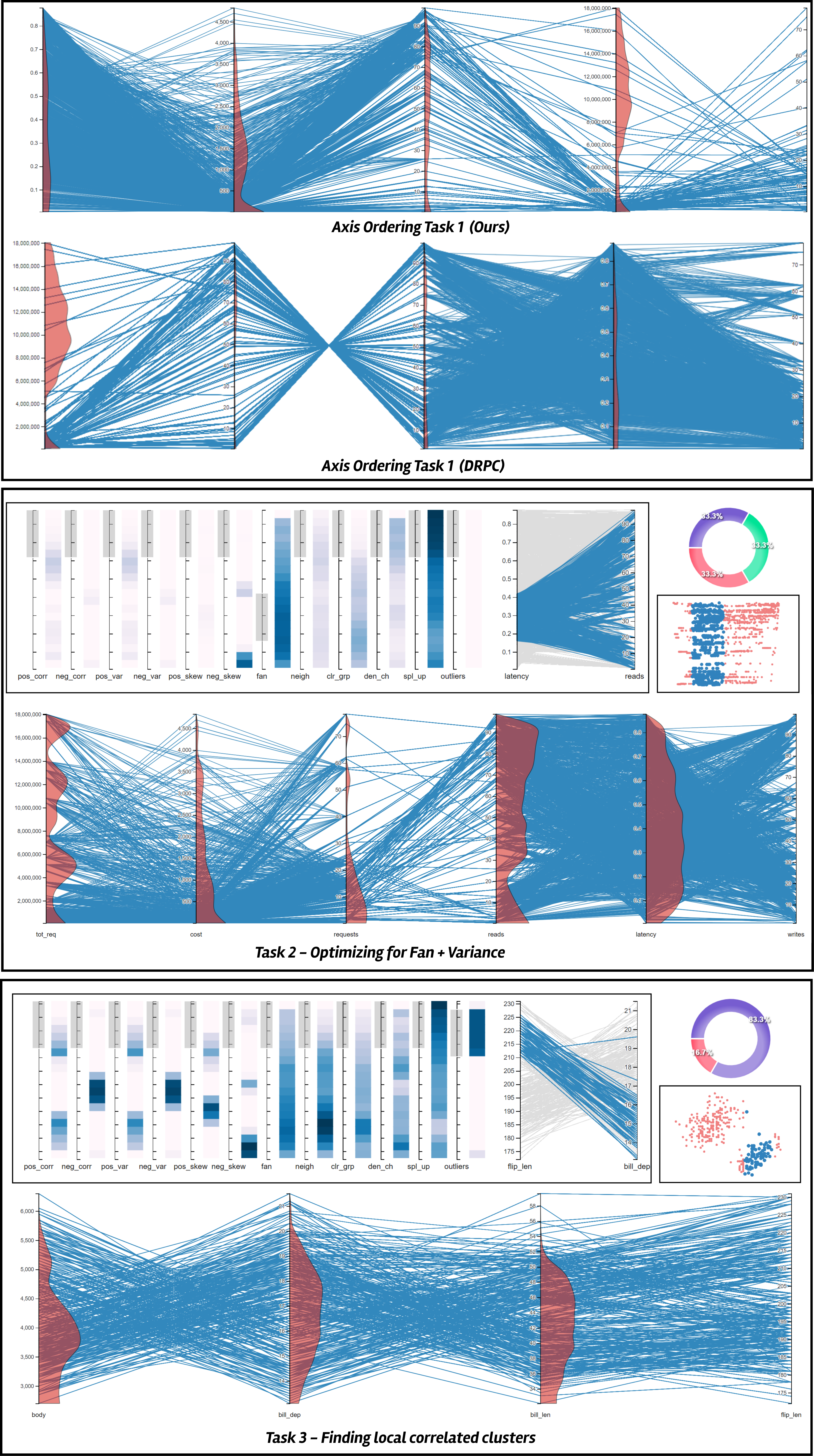}
 \vspace{-7mm}
 \caption{\textbf{(TOP)} Sample PCP orderings generated by a user
   during Task 1 of the user study (see
   Section~\ref{sss:user_study_tasks}).
 The results compare \name with the baseline tool
 DRPC~\cite{blumenschein_evaluating_2020}, showing that the user arranged
 the axes better using \name based on the area charts on the final PCP shown. \name shows a constant detection of the properties
 at all the axes.  In DRPC, users could only find high values along
 the first axis, while the other axes do not show any detection of
 the properties.  \textbf{(CENTER)} Axes ordering sample from a user
 performing Task 2 in \name.  The task was to optimize for \textit{Fan}
 and \textit{Variance}.  In the result, a clear fan structure is shown
 in the first three axes and higher variance is visible in the last
 two dimensions.  \textbf{(BOTTOM)} Axes ordering for user study Task
 3 as performed by one of the users on \name.  The task was to find
 local correlated clusters in the dataset.  The final ordering in the PCP
 shows highly negatively correlated clusters in all the axes as
 detected by the user.  High-resolution images are provided in the supplementary material.}
 \label{fig:task1_example}
\end{figure}

Employing a within-subject design, we created three tasks for this
user study to compare \name and DRPC on multiple aspects of PCP axes
reordering.  All the tasks required users to come up with a final PCP
axes reordering which they thought worked best.  For a fair comparison,
every task was related to correlation and clustering analysis, since it is
well supported in the baseline tool.  Based on the formative study
(Section~\ref{s:formative_study}), this setup covers general cases
that analysts encounter in data storytelling with PCPs.  Also, the tasks
were ordered in increasing levels of the detailing needed to generate the
final result.  The time and number of operations taken by the user to
generate final result were logged for every task.  To quantify the
goodness of final PCP axes ordering produced by the participants,
final scores were generated using the predefined properties for that
task (\eg as per Figure~\ref{fig:task1_example}, top).  Since the
localization is undefined for DRPC-based reorderings, an average of
all the localization levels supported in \name was used to score the
axes orderings obtained with DRPC. For a fair comparison, users were not allowed to see the final score through the area charts on the final axis orderings, as it was a direct proxy to the final score. 

\textbf{Task 1} focused on global, fully automated PCP axes
reordering use cases.  Users were asked to reorder the PCP axes
to best present \textit{correlated clusters} and to separate
\textit{outliers} in the data.  For the baseline, the task involved
experimenting with existing methods for
correlation~\cite{dasgupta_pargnostics_2010,
  berchtold_similarity_1998, artero2006enhanced} and
outliers~\cite{peng_clutter_2004} based PCP reordering.  \name,
can perform the task by choosing appropriate property weights on
the sidebar and looking at the corresponding generated PCPs.  The final
scores of the properties \textit{correlation and outliers} were
calculated by using \name to compare the goodness of generated PCP
orderings from
both tools (see Figure~\ref{fig:task1_example}, top).
This task evaluated \name for the speed and accuracy of the local property
detection algorithms it used.

\textbf{Task 2} focused on cluster identification with maximum
variability.  In this task, participants were asked to reorder the PCP
to show the maximum number of small clusters on one axis that spread on
the other axis (\ie fan-shaped clusters with high divergence).  This
scenario is common in general analytical tasks where the users seek to
find the behavior of some data sample with respect to a dependent
variable.  For example, in the computer systems domain a relationship
between \textit{``hard drive type''} and \textit{``total requests''}
follows this pattern because hard drives that can handle more requests
are preferred (see Figure~\ref{fig:task1_example}, center).  For final
evaluation, the scores of the properties \textit{fan} and
\textit{outliers} were used to compare the axes orderings.  This task
evaluated \name for HIL axes reordering efficiency.

\textbf{Task 3} focused on finding small regions of similar behavior
in the data.  Users were asked to reorder the PCPs to show the maximum
number of local structures in the data based on correlation (positive
or negative).  Some examples of this scenario include finding the
relationship between \textit{``miles per gallon''} and
\textit{``weight''} in the cars dataset.  Cars with higher weight
generally have low MPG and vice versa.  Building upon
task 2, this task emphasized finding local regions of interest
in the data.  Figure~\ref{fig:task1_example} (bottom) shows an example
user-generated PCP ordering for this task.  Final scores for the
participant reorderings were generated using the \textit{clear grouping},
and \textit{correlation} properties in \name.  This task evaluated
\name for multi-property HIL axes reordering.


\subsubsection{User Study Dataset}

We used two datasets for this study: a systems dataset for tasks 1 and
2 and the penguins~\cite{penguins} dataset for task 3.  The systems
dataset is larger and has more complicated patterns that are not
straightforward to visualize.  The penguins dataset, however, is less
complex with simpler-to-detect local patterns.

The \textbf{Systems dataset} was collected over a set of several
experiments run at our university to record the system performance for
a large number of configurations.  Currently, the dataset consists of
10 dimensions with a total of 100k configurations, giving the recorded
performance
values for each.  For the tasks to run in real-time on the baseline tool, we
sampled the dataset down to 2k rows and 6 dimensions, \textit{total requests},
\textit{cost}, \textit{requests per second}, \textit{number of reads},
\textit{latency}, and \textit{number of writes}.  This
allowed for a fair comparison of time and number of operations between
\name and the baseline (DRPC).  This dataset was chosen for the study
because the final PCP reordering results could be evaluated by systems
researchers.  Also, some of the user study participants were not from the
systems domain, allowing a fair comparison with no
preexisting knowledge of data features.

The \textbf{Penguins dataset}~\cite{penguins} is a
publicly available dataset with 2k rows and 6 features, containing
details about different species of penguins.  This dataset contains
simple local features, which better suited Task 3 because of time
constraints on the tasks.

\subsubsection{Procedure}

During the study, participants completed the above three tasks using
both DRPC and \name, one after another.
Prior to the tasks, participants were first introduced to both
systems.  Participants were allowed to play with the tools using some
pre-determined PCP reordering examples, independent of the user study
tasks.  After the users were comfortable using the tools, they were
introduced to the tasks sequentially.  Each task was limited to
20 minutes, and the time and number of operations were logged for each
user.

\subsubsection{Results}

\begin{figure}[tb]
 \centering
 \includegraphics[width=\columnwidth]{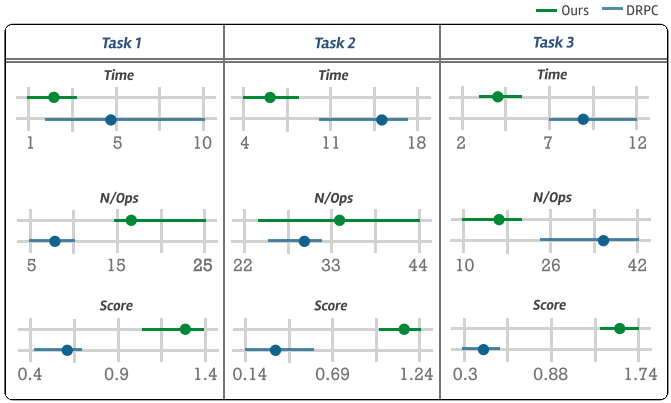}
 \vspace{-6mm}
 \caption{Results for time, number of operations (N/Ops), and scores
   for the three tasks used to compare \name with the baseline tool
   (DRPC).  The lines show the range of values
   obtained from the user study participants for each metric.  The dot
   on each line
   represents the distribution mean.  This user study evaluation shows
   that with \name, faster PCP axes reorderings can be generated with
   higher accuracy compared to the baseline.  Significant p-values, $<0.005$
   were obtained for all these experiments for the Wilcoxon test.  Details
   are provided in the supplementary material.
   }
 \label{fig:tasks_eval}
\end{figure}

Figure~\ref{fig:tasks_eval} summarizes the user study results obtained
from the participants' interactions for \name and DRPC.  We compared the
two tools based on three factors: \textit{time}, \textit{number of
  operations(N/Ops)}, and \textit{score} of the final PCP orderings
generated by the participants.
%

The results show that the average participant time for axes reordering
tasks was lower in \name compared to the baseline.  For task 1, the
average time on \name was 1.8 minutes compared to the 5.1 minutes on DRPC.
For task 2, the average time for \name was 5.3 minutes compared to 14.8
minutes on DRPC.  And for task 3, the average time for \name was 4.3
minutes compared to 9.1 minutes on DRPC.  There is a clear trend of
increasing time difference between the baseline and \name as the task
complexity increases.  This indicates the efficacy of \name in
faster PCP axes reordering.

For number of operations (N/Ops), our user study showed mixed results.
In task 1, the average N/Ops for \name were 18.5 compared to a lower
number, 7.7, for DRPC.  A similar pattern was seen during task 2 where
the average N/Ops for \name was 34.8 compared to 31 for DRPC.
However, in task 3, the average N/Ops in \name was lower: 15.8
compared to the 33.4 in DRPC.  The higher N/Ops for \name can be
attributed to the local view because the users validate their choices
through the local view interactions.

For the third set of result attributes, we compared the final scores
obtained from the PCP axes ordering finalized by the participants.
These PCP orderings were input to \name and the value of the final
area charts was compared.  \name had better results in all three
tasks compared to the baseline in this case.  The average score for
task 1 with \name was 1.13 compared to 0.62 for DRPC.  For task 2
this gap increased, with an average of 1.09 for \name and 0.32 for
DRPC.  For task 3, \name had an average score of 1.64 compared to 0.44
for DRPC.  This clearly shows the value of local properties
visualization supported in \name, which aids in better final PCP designs
compared to the baseline.

Comparing the numbers from the results, it is clear that even though
users end up finding higher N/Ops while working with \name, the
quality of the results is better and the time taken to perform the
axes reordering
is lower.  Each interaction in \name aids in better development
towards a final result, and the users reach the goal faster and with
higher accuracy.  Full results from the user study and the numbers for
each participant and tasks are provided in the supplementary material.

\begin{figure*}[ht]
 \centering
 \includegraphics[width=\textwidth]{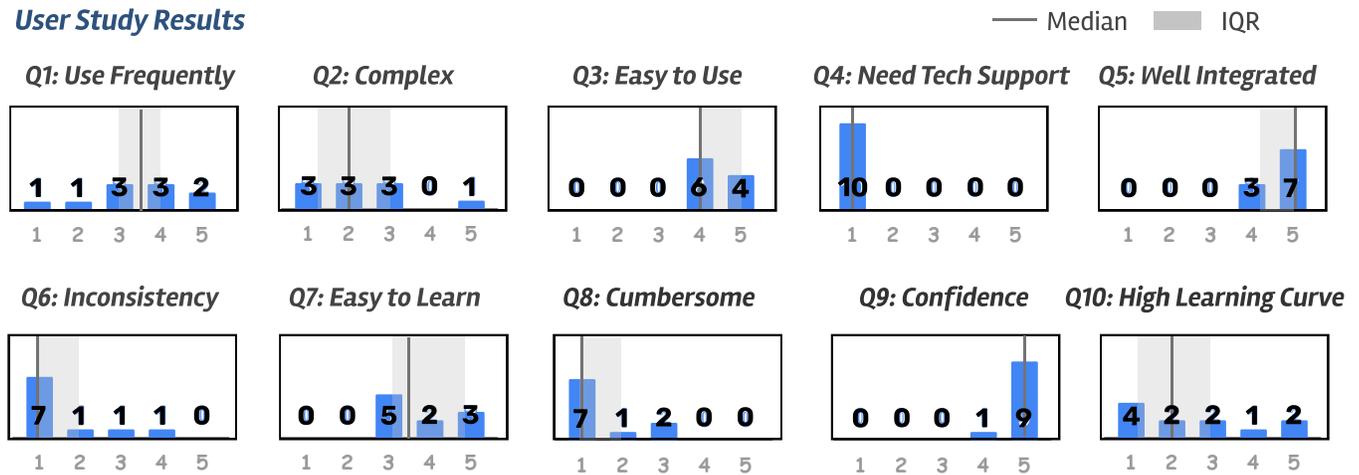}
 \vspace{-8mm}
 \caption{SUS~\cite{bangor2009determining} study questions and
   participant votes on a scale of 1 (Strongly Disagree) to 5
   (Agree).}
 \label{fig:sus_eval}
\end{figure*}

\begin{figure}[ht]
 \centering
 \includegraphics[width=\columnwidth]{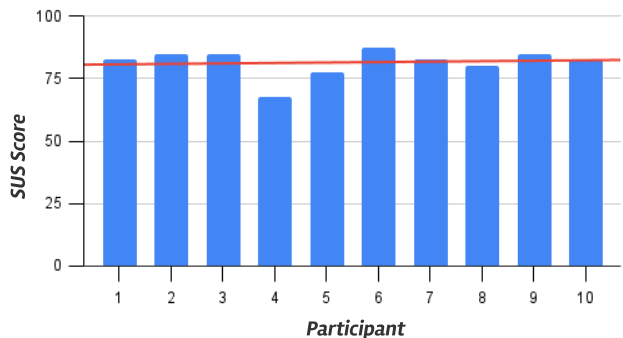}
 \vspace{-8mm}
 \caption{SUS~\cite{bangor2009determining} evaluation scores for \name
   obtained from the 10-user study participants, summarizing their
   experience.  \name received an average SUS score of \textbf{81.5},
   which is considerably higher than the baseline SUS score of 68 for
   design tools.}
 \label{fig:sus_eval_users}
\end{figure}

\subsection{System Usability Study}
\label{ss:sus}

After collecting the results from the user study tasks, every
participant was interviewed to collect feedback on the usability of
\name.  This study aimed at evaluating \name for multiple factors
based on the system usability scale
(SUS)~\cite{bangor2009determining}.  The SUS score is an industry
standard for quantifying the usability of any visual analytics tool
through a series of questions that focus on evaluating
different aspects of the tool and are sequenced alternatively to focus
on positive and negative aspects.  Participants are required to
answer each question on a 5-point Likert scale that ranges from
Strongly Disagree (1) to Strongly Agree (5). The question sequence is altered during the interview to collect positive and negative feedback about the tool. 
Figure~\ref{fig:sus_eval} shows the 10 SUS questions used to evaluate
\name.  The whole interview lasted about 10 minutes for each
participant.

\subsubsection{Results}

As shown in Figure~\ref{fig:sus_eval_users}, \name received a
positive usability feedback from a majority of the participants.
Overall, based on participant ratings for the 10 SUS questions, \name
received an average SUS score of 81.5.  Since the baseline SUS
standard is 68~\cite{bangor2009determining}, the SUS evaluation
results for \name show that our system is highly adaptable and has
great usability in this domain.
We also separately evaluated the SUS scores for experts and
non-experts in this study.  The expert SUS scores averaged 84 while
the non-expert scores averaged 79, both being higher than the baseline
score.  Detailed scores and analyses for each participant are provided
in the supplementary material.

Further analyzing the scores from the participants for
individual SUS questions (see Figure~\ref{fig:sus_eval}), the
questions receiving the most unanimous high votes included (Q4) \textit{no
  need for tech support} and (Q9) \textit{confidence}.  These results
indicate that \name's design is easy to use and users are highly
confident in the generated results, complementing the fundamental
principle of explainability for developing the tool.
Questions receiving the second-highest vote counts included (Q3) \textit{Easy
  to use}, (Q5) \textit{Well integrated}, and (Q6)
\textit{Consistency}.  These results show that \name
aligns well with the goals of real-time and accurate implementation of the
PCP reordering schemes.  Overall, the average IQR
of votes across all
the questions was 1.25, which is considered low (good)
%
%
on a 5 point scale.  This shows that both experts and non-experts had a
similar experience with \name, and hence the tool is consistent across
users.

Besides the majority of positive feedback, a few questions received mixed
reviews during the study.  (Q1) \textit{Use frequently}, had two low
votes from experts, which we attribute to the limitations of PCPs with very
high-dimensional datasets (discussed further in
Section~\ref{ss:user_interviews}).  Also, (Q10) \textit{High learning
  curve}, had three low votes from non-experts participants.
To further evaluate these issues, we conducted a
 detailed interview session, discussed next in the follow up text.

\subsection{User Interviews}
\label{ss:user_interviews}

Besides the general SUS interviews, we separately collected detailed
feedback from two study participants with the lowest SUS scores: an
expert (E1) and a
non-expert (E2).  These interviews helped us
gain further insights into user experience with \name, its
limitations, and potential areas of improvement.  Some of the interview results
and user comments are discussed below.  We have categorized the user
comments based on their experience with \name. Three extra question discussions are provided in the supplementary material. 



\textbf{Ease of exploration.} Both participants agreed that the local view
and heatmap reordering were useful in exploring local regions in the
data.  E1 suggested an improvement in their comment:
\textit{``Sometimes we want to intentionally ignore a property, for
  example, ignore outliers while optimizing for other properties.  In
  such cases, negative weights for some properties would be
  helpful.''}

\textbf{Use on regular basis.} Both users gave positive reviews on the
usability of \name, with a few suggestions for improvement.
\textit{P1} commented: \textit{``I regularly come across the issue
  with PCPs when the dimensionality of data increases.  Very high
  dimensional datasets are hard to fit with PCPs.  Maybe involving
  dimension reduction techniques to shrink data dimensionality before
  analysis with \name will be a good addition.''}
E2 further suggested: \textit{``I really found \name very useful for
  data storytelling with PCPs. Every axis ordering presents a different story in the data and \name makes it easy for us to get the right story out with PCPs. I suggest adding similar support for
  tools that work with categorical data will be a good next step.
  Parallel Sets can be extended with this local visualization of
  properties too.''}



\textbf{Ease of usage and creativity.} Both participants mentioned
that they scored \name high on this criterion.  E1 commented:
\textit{``It is very easy to understand line patterns and local
  properties in my data with \name.  Since now we have implemented
  real-time detection schemes for several line patterns, it will be
  interesting to extend this work with machine learning.  Maybe we can
  create a small dataset of detected patterns with \name and use it to
  train a supervised model.''}
E2 also commented about the multi-property axes reordering:
\textit{``This was the first time I experimented with multi-property
  reordering in PCPs.  It was very easy to see which was a dominating
  property for a particular axes pair through \name.}


\section{Conclusion}



Axes reordering plays a crucial role in HD data storytelling with
parallel coordinate plots.
In this work, we present \name, an all-in-one, real-time, human-in-the-loop,
localized PCP axes reordering framework.  Our framework implements
real-time detection algorithms for the 12 most common data analytical
properties occurring in PCPs as shown in past
research~\cite{blumenschein_evaluating_2020}.  This enables \name to
extend the pattern detection from a full axis scale to a more
localized approach.  With \name, we can detect several data patterns
on a local level and visualize their behavior
across all dimensions.  These local-detection schemes allow
further advancements in PCP axes reordering techniques.

\name supports two types of axes reordering schemes: fully automated and
human-in-the-loop (HIL).  In the fully automated method,
users can control the localization level in \name and
calculate pairwise data dimension scores.  Using these scores, the
final ordering can be generated using a
traveling salesman~\cite{flood_traveling-salesman_1956} algorithm.
Conversely, in the HIL paradigm, \name supports local views of
pairwise dimensions, highlighting zones of activity for the
user-selected properties in the data.  This information can be used to
manually generate the ordering of axes as desired for different
use cases.  And when every data ordering in a PCP presents
a different story, no single axes ordering algorithm can suit all
usage scenarios.  \name aims to assist the users in presenting
\emph{their} story through PCPs with high accuracy and confidence.

We learned several important lessons while designing \name.  Our
initial discussions with domain experts during the formative study
were decisive in pinning down the main design components.  After all
the tasks and contributions were formulated with the experts' help, it
was easier to design the visual interface for \name with all the
components.  Primarily, we realized that adding a human in the loop is
important as a means of allowing users to infuse their own domain
knowledge into the process.

\paragraph{Future work.}
Beyond \name's effectiveness, there are
aspects that can be improved.  Our interviews with the user-study
participants pointed out several features and directions for future work.
First, adding support for negative-weight properties is useful if
the user wants to actively ignore a particular
property. Also, scalability is an issue with \name when the data dimensionality goes beyond 15 dimensions. The matrix view gets cluttered and calculation of metrics on window sizes smaller than 30\% becomes slower. To overcome this, data filtering and dimension reduction techniques can be incorporated into \name.  
We would also like to explore different metric combination measures besides weighted sum. Since weighted sum has known limitations, for e.g. the distribution of the solution space is not uniform~\cite{das1997closer}, we would like to explore other distance measures to combine the metrics. 
Additionally, we plan to add a history feature to \name;
this  will allow users to backtrack the process of
ordering the axes through the heatmap.
One of the user study participants also suggested additional support
for high-dimensional datasets using dimension-reduction techniques.
We can reduce the dimensionality of the data to  make it easier to
represent information through a PCP and allow faster property
calculation.
More advanced additions to \name include storytelling with natural
language processing.  The idea is that we can combine the properties
in PCPs to tell a common story that they represent in the data.  In this
way, every PCP axes ordering can be linked to a data story that can be
presented in a few sentences. Also, it'll be interesting to know if pre-classification of data records locally would help in further improving the axis ordering techniques. These features are not yet supported,
and we  continue to design and develop  \name.
Finally, we plan to deploy \name for real users to collect in-depth
user feedback in a longitudinal study.

\section{Acknowledgements}
We would like to thank the anonymous VIS 2022 reviewers for their valuable comments. This work was partially funded by NSF grants CNS 1900706, IIS 1527200 and 1941613, and NSF SBIR contract 1926949.


\bibliographystyle{abbrv-doi}
\bibliography{references}
\end{document}